# *Reliability and validity of TMS-EEG biomarkers*

Sara Parmigiani[1,2]
Jessica M. Ross[2,1]
Christopher Cline[1,2]
Christopher Minasi[1,2]
Juha Gogulski[1,2,3]
Corey J Keller[1,2,δ]

[1]Department of Psychiatry and Behavioral Sciences,
Stanford University Medical Center, Stanford, CA, 94305, USA

[2]Veterans Affairs Palo Alto Healthcare System, and the Sierra Pacific Mental Illness, Research, Education, and Clinical Center (MIRECC), Palo Alto, CA, 94394, USA

[3]Department of Clinical Neurophysiology, HUS Diagnostic Center, Clinical Neurosciences, Helsinki University Hospital and University of Helsinki, Helsinki, FI-00029 HUS, Finland

[δ]Correspondence:
Corey Keller, MD, PhD
Stanford University
Department of Psychiatry and Behavioral Sciences
401 Quarry Road
Stanford, CA 94305-5797
Email: ckeller1@stanford.edu
Phone: +1 8025786292

To be submitted to: Biological Psychiatry: CNNI

Number of pages: 26
Number of figures: 2
Number of tables: 0
Abstract word count (250): 250
Total word count (4000): 3990

**Running title:** Reliability of TMS-EEG

**Keywords:** TMS, EEG, TEP, reliability, validity, TMS-EEG

**Funding:** This work was supported by R01MH126639, R01MH129018, and the Burroughs Wellcome Fund Career Award for Medical Scientists. JR was supported by the Department of Veterans Affairs Office of Academic Affiliations Advanced Fellowship Program in Mental Illness Research and Treatment, the Medical Research Service of the Veterans Affairs Palo Alto Health Care System, and the Department of Veterans Affairs Sierra-Pacific Data Science Fellowship. JG was supported by personal grants from Orion Research Foundation and Finnish Medical Foundation.

**Abstract**

Noninvasive brain stimulation and neuroimaging have revolutionized human neuroscience, with a multitude of applications including diagnostic subtyping, treatment optimization, and relapse prediction. It is therefore particularly relevant to identify robust and clinically valuable brain *biomarkers* linking symptoms to their underlying neural mechanisms. Brain biomarkers must be reproducible (*i.e.*, have *internal reliability)* across similar experiments within a laboratory and be generalizable (*i.e.*, have *external reliability*) across experimental setups, laboratories, brain regions, and disease states. However, *reliability* (*internal* and *external*) is not alone sufficient; biomarkers also must have *validity*. *Validity* describes closeness to a true measure of the underlying neural signal or disease state. We propose that these two metrics, *reliability and validity*, should be evaluated and optimized before any biomarker is used to inform treatment decisions. Here, we discuss these metrics with respect to causal brain connectivity biomarkers from coupling transcranial magnetic stimulation (TMS) with electroencephalography (EEG). We discuss controversies around TMS-EEG stemming from the multiple large off-target components (noise) and relatively weak genuine brain responses (signal), as is unfortunately often the case with human neuroscience. We review the current state of TMS-EEG recordings, which consist of a mix of reliable noise and unreliable signal*.* We describe methods for evaluating TMS-EEG biomarkers, including how to assess *internal* and *external reliability* across facilities, cognitive states, brain networks, and disorders, and how to *validate* these biomarkers using invasive neural recordings or treatment response. We provide recommendations to increase *reliability* and *validity,* discuss lessons learned*,* and suggest future directions for the field.

## 1. *Reliability* and *validity* in neuroimaging

Noninvasive brain imaging and stimulation has revolutionized human neuroscience over the past thirty years. Many tools exist to image brain activity including functional MRI (fMRI), functional near-infrared spectroscopy (fNIRS), magnetoencephalography (MEG), and electroencephalography (EEG). However, each modality has different limitations in terms of cost and complexity (fMRI, MEG), limited temporal resolution (fMRI, fNIRS), and limited spatial resolution (fNIRS, MEG, EEG). Pairing these imaging techniques with noninvasive brain stimulation can provide additional power by enabling the study of responses to perturbation in focal regions of cortex and connected networks. These causal methods include using transcranial direct and alternating current stimulation (tDCS and tACS, respectively) and transcranial magnetic stimulation (TMS), and have been critical in accumulating knowledge of and developing novel treatments for neurological and psychiatric disorders. To use noninvasive brain stimulation techniques as part of a biomarker, stimulation responses must be quantified with a metric related to normal neurophysiology, pathophysiology, or responses to an exposure or intervention (1,2). TMS effects on the motor and visual cortices can be captured with corticospinal (electromyography) and visual perception readouts, but TMS to other regions may be best captured using neuroimaging The combination of TMS with EEG is particularly promising due to the comparable temporal resolution between TMS and EEG and the possibility of improving EEG's poor spatial specificity using the greater spatial specificity of TMS itself.

Given the limitations of each of these noninvasive tools, it is critical that the readouts of stimulation-imaging methods are shown to be stable and repeatable (*reliability*) and to measure the underlying neural processes of interest (*validity*). Specifically, these metrics need to have *internal reliability* across similar experiments within a laboratory, answering the question: *'how well in the lab and for this experimental setup can the study be reproduced?'* They also need to have *external reliability,* answering the question: *'how well in other laboratories or clinical environments, with different experimental setup and where operators may be differently trained, can the study yield consistent results?'* There are many instances in which a metric can have *validity* but not *reliability* or *reliability* but not *validity* (Figure 1). We propose that it is necessary to critically evaluate and optimize both the *reliability* and *validity* of any brain biomarker of interest, particularly prior to implementation in aiding the diagnosis or treatment of any neurological or psychiatric disorder.

All too often, there is an understandable desire to immediately use recently-discovered human noninvasive tools rather than to first send them through months of rigorous testing. Rigorous testing is time-consuming and often less immediately impactful and application of a novel tool to study a specific brain circuit or disorder can yield high impact publications, open up future lines of investigation, and be immediately translatable to the clinic. However, if a tool has either low *reliability* or *validity*, clinical application should not proceed until the tool is deemed sufficient by the scientific community. Unfortunately, and particularly for tools not regulated by the FDA, there is no such rigorous statistical barrier to mainstream use. Instead, rigorous testing is often performed after years of application research (3,4). A lack of rigor in biomarker development can in the end impede scientific progress and cast doubt on the measurement tools, over time weakening the scientific community's view on clinical noninvasive human neuroscience (5).

We propose that prior to clinical use, noninvasive human neuroscientific metrics must be rigorously evaluated and deemed to have high *reliability* and *validity*. The field of *psychometrics* is dedicated to the evaluation of scientific metrics of psychological properties, but is most often applied to clinical and behavioral assessments and not regularly to novel neuroimaging biomarkers of interest (6,7). Thus, we must develop these psychometric-like rigorous testing algorithms for novel noninvasive brain measurements prior to clinical use. Below we outline this process for one relatively new noninvasive brain measurement tool, TMS-EEG, and describe how to rigorously assess and optimize TMS-EEG biomarkers. We hope that other research endeavors follow suit in rethinking, reevaluating, and improving biomarkers from their measurement tool of choice.

## 2. Motor-evoked potentials (MEPs): A gold standard

While noninvasive imaging techniques (fMRI, MEG, EEG) are powerful metrics of neural activity, they lack a causal component. Causal techniques such as TMS involve perturbing brain activity and measuring the consequent response. Initial TMS studies focused on stimulation of the primary motor cortex (M1) (1–4)(12), where much of our knowledge about the physiological effects of TMS in humans is derived (13–15). Thanks to the somatotopic organization of the M1, a muscle of interest can be activated by applying TMS to a specific portion of the M1 (16). Supra-threshold single TMS pulses over M1 elicit a strong electromyography (EMG) response over skeletal muscle, termed the *motor evoked potential* or MEP. The MEP is linearly correlated with the number of activated corticospinal neurons (13) and thus considered *valid* to track cortico-spinal excitability. Based on many studies, the MEP has been shown to be highly *reliable*, stable across time within a laboratory (*internal reliability*, (17–21)), and generalizable across laboratories (*external reliability;* (10,22–25)), partially due to its high signal-to-noise (>1000 uV responses with low levels of noise). Of course, some theoretical assumptions and approximations are required, but all things considered, the MEP and its features (*e.g.*, latency, amplitude, morphology) have high *internal* and *external reliability*, making it a *valid* biomarker to track excitability of the corticospinal tract. As such, it is in widespread use both in research and clinical practice (26–28). Unfortunately, MEPs can only probe the corticospinal tract, and as a result other tools including TMS-EEG are needed to explore causal relationships across other brain regions.

## 3. TMS-evoked potentials (TEPs)

Although relatively new, TMS-EEG is a powerful noninvasive neuroimaging tool with strong *internal reliability* (29). However, more rigorous *external reliability* and *validity* assessment with subsequent optimization is needed. TMS-evoked potentials (TEPs) are a result of coupling single pulses of TMS with scalp EEG recording. TEPs were first described in 1997 (30) with a more mainstream audience in 2005 (31). In contrast to MEPs which probe the central and peripheral nervous system response together, TEPs measure the central nervous system response to single pulses of TMS. TEPs consist of a multiphasic response lasting ~500 ms (32). Initial reports described high *internal reliability* (Figure 2; see (29)), demonstrating that TEPs can be repeatable after a week and are sensitive to changes in stimulation amplitude, site, and angle (29). Subsequently, TEPs were largely viewed as plug-and-play and have since been applied to various cognitive states and connectivity features (33,34), brain disorders (35,36), stimulation sites (37), stimulator devices (38), amplifier types (38), and EEG recording setups (39).

However, the assumed *external reliability* began to be questioned. Initial studies were performed by stimulating medial structures with minimal muscle artifact (29,31,40), which may explain the high reports of *internal reliability*. It was eventually determined that 1) stimulation to lateral brain regions elicits strong early muscle artifacts that can confound the TEP (41,42), and 2) the sensation and auditory click from a TMS pulse results in non-specific sensory responses (43). As a result, controversies ensued when TEPs across laboratories elicited different levels of *reliability* of the TEP. In recent years, multiple groups have re-evaluated the *internal reliability* of TEPs (44–46). Previously demonstrated substantial *internal reliability* for later components (>50 ms) of the TEP are now known to consist partially of sensory non-specific off-target effects (43), and weaker *internal reliability* of earlier components (<50 ms) of the TEP are thought to reflect more *valid* components of local cortical excitability (47). Discussion of the *reliability* and *validity* of these early and late components of the TEP collided when similar TEP morphologies were reported after single pulses of sham TMS (5) and single pulses of TMS to the shoulder (48). In response, a heated discussion ensued within the TMS-EEG scientific community questioning most appropriate methods for obtaining a true brain TEP recording (Figure 3, (46)). This discussion highlighted the need for 1) rigorous assessment with each new TMS coil, location, angle, and intensity probed, as well as 2) standardization across the TMS-EEG community with respect to *reliability* and *validity* assessments.

To move towards robust measures of causal brain excitability, particularly for extended use to clinical applications, we evaluate in the sections below the off-target and intended neural signals present in the TEP, discuss these in reference to *reliability* and *validity* metrics, and outline methods to assess and enhance these metrics.

### 4. Off-target components (noise) in the TEP: reliable but not valid

The largest amplitude components in the TEP are usually *reliable* but not *valid*. Because the TEP is the result of single TMS pulses averaged across multiple trials (49,50), it is an aggregate of all electrical signals measurable using EEG. Contributing sources to the TEP have overlapping time courses and are thus susceptible to net displacement effects of individual waves (*i.e.*, constructive and destructive interference). In addition to multiple central neural sources (30,30,37,51), contributions to this aggregate waveform also include non-neural sources (52–56) and peripherally-evoked neural sources unrelated to the direct effects of but time-locked and in response to TMS (5,48,57–60). TMS-evoked central neural sources will be referred to hereafter as *signal,* and non-neural sources, peripherally-evoked neural sources, and off-target neural sources will be referred to as *noise*.

Myriad sources of *noise* confound interpretation of the TEP. Non-neural contributions include 1) pulse artifact from current flow through the coil, 2) recharge artifact from the capacitor between discharges, 3) decay artifact from changes in capacitance between the electrodes and gel as well as between the gel and the scalp, 4) electrode noise from poor contact with the scalp or conductive gel drying or leaking, 5) line noise from equipment in the room (60Hz or 50Hz), 6) evoked and continuous muscle activity, 7) electrocardiogram (EKG), 8) eye blinks, eye movements, microsaccades, and 9) movement of the electrodes (often where the coil is resting on the cap) (53,56,61–63). Peripherally-evoked contributions to the TEP include 1) somatosensory-evoked potentials from the sensation of the TMS pulse on peripheral nerves and 2) auditory-evoked potentials from the auditory click of the TMS pulse (5,43,47,48,57,59,64). Off-target central neural sources can also occur due to 1) inhibitory mechanisms, including those associated with visual field changes from eye blinks, eye movements and microsaccades (65,66), 2) fluctuations in brain state such as awakeness and attention (59,65,67), and 3) pain-related responses (68). Best approaches for handling these many off-target contributions are the topics of numerous other publications and methodological debates (4,56,64,69–72). However, it should be emphasized that these non-neural, peripherally-evoked, and central off-target components of the TEP are often *reliable,* with large amplitudes and long durations (69,73), and therefore should be carefully considered and minimized using a combination of techniques during data collection (31,43,47,57,74,75) and during post-processing (4,48,56,64,73). In summary, there are numerous sources of *reliable* and easily recognizable large amplitude *noise* contributions to the TEP that need to be accounted for when designing a study, minimized during data collection, and if needed removed during data analysis to uncover the effective *signal* of interest.

## 5. True TEP components (signal): valid but unreliable

While obtaining true neural TEP *signal* is challenging due to the multitude of large magnitude off-target recordings often also present in the TEP (*noise*; see Section 4 and (42,76)), the observation of *valid* components of the TEP from locally-excited brain tissue is possible and can provide important insights into brain physiology and pathology. When TMS is applied to brain regions where off-target noise confounds are minimal, there is strong evidence for the neural basis (*validity*) of TEPs, especially in the early (<50 ms) components (47). For example, in primary motor cortex (M1), the MEP correlates with the amplitude of the early 15-30 ms response of the TEP (77,78), suggesting that early TEP components reflect cortical excitability. Moreover, local high-frequency TEPs evoked by M1 stimulation and paired-pulse MEPs (conditioned MEPs) share a similar time course, suggesting that they may index similar types of cortical activity (79). Furthermore, TMS to the prefrontal cortex (PFC) produces TEP peaks (80) with amplitudes (81) and latencies (82) that correlate with M1 TEP amplitudes. However, TEPs evoked by PFC and M1 stimulation may reflect different underlying neural mechanisms: early TEP peaks (<50 ms) following left PFC stimulation are insensitive to excitatory NMDA receptor blockade (83), whereas early M1 TEP peaks reflect a balance of GABAergic inhibition and glutamatergic excitation (84). TEPs have been examined in various neurological and psychiatric disorders (for review see (50)), but these studies have largely focused on later (>100 ms) components of the TEP, where the peripherally-evoked off-target effects are known to be present and have large amplitude, confounding the *validity* of these biomarkers. In summary, although there is evidence suggesting *validity* of the early components of the TEPs, the off-target neural sources and non-neural artifacts surrounding these early components hinders *reliability* and exploration as potential biomarkers of diagnosis and treatment outcome. Thus, increasing the amplitude of these early TEPs and removing artifact and other off-target effects is a critical next step to boost both *reliability* and *validity* of the TEP.

## 6. Evaluation of TEP reliability

A *reliable* TMS-EEG biomarker should be stable across time (85,86) when there is no change in the indicated disease process, and should be sensitive to change when fundamental parameters including instrument, recording set-up, cortical area stimulated, or intensity of stimulation are varied (29). *Reliability* of a biomarker can be evaluated by: (i) measuring the same metric over time within a laboratory (minutes (36), hours (85), or weeks (25,26,76) in a test-retest fashion (25,28,69,74)); (ii) repeated quantification across different instruments or in different laboratories (44,48,68); or (iii) against different preprocessing and analytic pipelines (4,61,64). Described in more detail below, we find that *internal reliability* can be high when using certain set-ups (26) and when applying TMS to medial brain regions with minimal muscle activation. *Reliability* is reduced with the presence of *noise* sources, including when more lateral sites with larger muscle activations are stimulated. Methods to remove sources of *noise* and boost *reliability* include experimentally minimizing sensory contributions online (64) and offline (56), and should be considered in reference to *reliability* metrics.

## 7. Boosting TEP *reliability* during data collection

With a thorough understanding of the factors contributing to *noise* and *signal* in the TEP, various methodological approaches during assessment can be employed to boost the signal to noise ratio (SNR) and improve *reliability*. EEG amplifiers with specialized features (high sampling rate, high dynamic range, slew-rate limiting, and/or sample-and-hold circuitry) can reduce the effect of the primary electromagnetic pulse artifact on the recorded EEG signals (88–90). Delaying stimulator capacitor recharge until after the TEP time period of interest can prevent recharge-related electrical noise from masking relevant EEG signals (91). Minimizing EEG electrode impedance (<5 kOhms) can reduce decay artifact caused by charge buildup at the electrode-gel-skin interface (92). Active electrodes can reduce sensitivity to environmental electrical noise such as 50 or 60 Hz line noise when compared with passive electrodes (93). A thin layer of foam placed between the coil and scalp may reduce artifacts related to bone conduction of the TMS sound (75). Passive noise reduction with earmuffs (43,47) and active noise masking (57,75) minimize auditory sensation and saliency of pulses. In combination with noise reduction and masking, alterations in pulse timing can further reduce off-target EEG components related to sensation and saliency (43). When feasible, rearranging electrode wires can minimize TMS-induced electrical artifact in electrodes near the site of stimulation (39).

Undesirable *noise* and desirable *signal* can be highly sensitive to stimulation location, coil angle, and intensity. By quantifying some aspects of *noise* and *signal* in real-time, it is feasible to individualize stimulation parameters (location, angle, intensity) to directly maximize the SNR of specific TEP features (29,94). Such real-time individualization approaches hold great promise for improving the *reliability* of TEP-based biomarkers by minimizing off-target effects and maximizing local brain responses. These approaches may also improve the *validity* of TEP-based biomarkers by effectively targeting stimulation to more relevant brain circuits with greater fidelity.

## 8. Evaluation of TEP *validity* using noninvasive tools

As stated above, early components (<50 ms) of the TEP may represent *valid* metrics of local cortical activation but are currently confounded by multiple sources of *noise*, which in turn reduces both *internal* and *external reliability.* The currently low accuracy of the early TEP with potential for high *validity* is deeply influenced by SNR, making an otherwise valid biomarker invalid (4). In the TEP, underlying signals of local activation can be masked by *noise* (95). It is worth noting that off-target sensory responses and true and valid TEP may not be independent and linearly separable phenomena. Thus, standard methods to remove these sensory responses and non-neural artifacts from genuine brain responses (96) as well as statistical comparisons between real and sham TMS (47,67,75,97–99) may be called into question and reveal the intrinsic challenge about the nature of TEP analysis. These limitations do not necessarily imply that the early TEP cannot be a valid biomarker, but rather that it requires further investigation, mechanistic understanding, sharing of data and protocols, and optimization of experimental setup to maximize SNR. Indeed, if we consider that *validity* describes how well a tool is sampling the desired physiology, in terms of local excitability, data gathered in the last few decades tells a promising while not complete story of the *possible validity* of early (<50 ms) components of the TEP. Indeed, early TEPs correlate with corticospinal tract excitability (78,100), are reduced in schizophrenia (101,102), mimics characteristic slow waves in NREM sleep, deep sedation, and disorders of consciousness (31,40,103–105), and are modulated in recovery from injury or stroke (106,107). To definitively link TEPs to their underlying neural correlates, however, intracranial brain recordings show promise.

## 9. Evaluation of TEP *validity* using invasive brain recordings

To evaluate the *validity* of a tool, in addition to linking novel neuroimaging metrics with well-known noninvasive measures, it is also important to link to intracranial neurophysiology. Stated another way, using insights gained from intracranial recordings as the 'ground truth' may be valuable to inform noninvasive TMS studies and establish the validity of TEP components. Intracranial neurophysiology can be valuable in a number of ways, including 1) TMS-evoked intracranial electrophysiology, 2) electrical stimulation-evoked electrophysiology, and 3) simultaneous invasive and noninvasive brain recordings. First, recent novel work combining *TMS with intracranial brain recordings* suggests that single pulses of TMS modulates both local and downstream brain circuitry that can be captured in the TMS-evoked intracranial response (108). Second, examining the intracranial neural responses to *intracranial electrical stimulation* can also provide valuable 'ground truth' information (109–113). As intracranial electrical stimulation is not perceived, the ground truth cortico-cortical response profile to electrical stimulation without sensory confounds can be compared to the noninvasive localization and morphology of TEPs. In this manner, one can identify and isolate TMS-evoked sensory responses from grounded TMS-evoked neural signals that are consistent with intracranial recordings. Finally, in contrast to the previous non-time-locked evaluations of noninvasive and intracranial measurements, a novel combination of simultaneous intracranial and EEG scalp recording after stimulation (12) can provide direct assessment of the neural correlates of noninvasive biomarkers. Caveats with this intracranial approach are that the patient population is currently limited to epileptic surgery patients and that there are physiological and artifactual differences between electrical and magnetic stimulation. In summary, pairing noninvasive and invasive brain stimulation with intracranial recording methods have great potential for improving the *validity* of TEPs and other novel neuroimaging metrics.

## 10. Boosting TEP *validity* using offline analytic methods

In addition to online optimization during data collection, offline analytical methods can be employed to remove *noise* and thus boost the SNR and *validity* of the TEP. There are many preprocessing pipelines that have been developed for this purpose (70–72,114), all of which involve a similar set of preprocessing steps: removal of 1) large TMS pulse artifacts, 2) line noise, 3) other known non-neural artifacts such as from eye movement and blinks, and 4) off-target neural sources (64), such as somatosensory and auditory activations. Removal of this *noise* is approached mainly using a combination of interpolation, filters, and independent component analysis (ICA) based methods (69,115). Interpolation is useful for artifacts with known timing and overwhelmingly high amplitude, such as the primary TMS pulse artifact, but inherently causes loss of data in the interpolated time range. This is why this technique is only typically used when the artifact overwhelms the overlapping EEG data. Filters break the signal into its spectral components and remove those outside of the specified range to effectively remove line noise, drift, and high-frequency muscle and environmental electrical noise. ICA is a blind source separation technique typically used for removal of eye-blink and saccade artifacts, and by many groups for removal of other off-target activations. ICA is used to decompose the EEG into independent sources of activity that are linearly mixed and is followed with identification of artifactual sources using known spatial and temporal characteristics and removal (69). This process can either be done manually or automatically using machine learning classification algorithms (72,116). In summary, offline removal of known artifacts in the TEP can boost SNR and enhance *validity* of the TEP.

After removing known *noise*, several further decisions regarding how to quantify the TEP are necessary, each of which can affect its *validity*. The TEP waveform is a complex multi-phasic response and quantifying this response is not straightforward. Many decisions are necessary including which peaks to quantify (P30, N45, P60, N100, and/or P200), how to quantify a peak (peak amplitude, peak-to peak-difference, area under the curve, latency, etc.), which electrodes to analyze (single electrode, multiple electrodes, sensor or source space), and whether to evaluate in time or frequency space (37,78). All of these decisions can greatly influence the *validity* of the TEP, since each metric extracts different information encoded in the TEP signal. For example, in assessing TEP change based on late TEP peaks (N100, P200), one might find more *reliable* (yet less *valid*) change since later TEP peaks are significantly composed of auditory-evoked potentials caused by the clicking sound produced by the TMS machine (5). Thus, in order to ask targeted questions regarding how the brain responds to the TMS probing of targeted neural circuits, it is essential that the metrics used to quantify TEP signals during analysis encode targeted and *valid* neural responses. Basic statistical techniques comparing the TEP after real and sham rTMS have been utilized to determine whether certain features, such as the late TEP peaks (98), represent *valid* measures of targeted neural activity. However, as there are nearly an unlimited number of features to extract from the TEP, more sophisticated statistical techniques (such as machine learning methods) could be helpful for data-driven approaches to making these decisions and should be further explored in this context. In summary, careful consideration of offline analytic decisions and incorporation of more modern data-driven approaches can greatly influence the *validity* of TEPs.

## 11. Recommendations and future directions

In this paper, we evaluate TMS-EEG, a powerful causal neuroimaging tool, to describe and discuss the concepts of *reliability* and *validity* of potential biomarkers. The TMS field started with a 'gold standard' MEP approach that was highly *reliable* (both *internally* and *externally*) and *valid*. As the field incorporated TMS-EEG approaches to evaluate cortico-cortical interactions outside of the motor cortex, initial groups reported high *internal reliability*. However, as TMS-EEG was adopted across labs and translated to different brain regions and experimental setups, *external reliability* and *validity* were both reduced. We speculate that the trajectory of investigating neurophysiological responses to TMS, from initial high *internal reliability* and *validity* to reduced *external reliability* and *validity*, is all-too-common in noninvasive neuroimaging. We hope that by adopting the recommended approach outlined below, this trajectory can be modified to produce tools with high *reliability* and *validity*.

To help guide further research into the *reliability* and *validity* of novel and established neuroimaging techniques, we provide some general recommendations:

- Although *reliability* and *validity* are related, a biomarker can have high *reliability* but low *validity* or high *validity* but low *reliability* (Figure 1). We suggest that careful consideration of both are necessary before implementing a biomarker to make diagnostic or treatment decisions.

- *Internal reliability* can be enhanced by optimizing the SNR during data collection. Hardware optimization, real-time search for optimal stimulation parameters, and closed-loop methodologies can enormously reduce *noise* and increase *signal* strength.

- Although initial reports of *internal reliability* may be high, we recommend immediate or very early assessment of *external reliability* by collaborating closely with one or more external labs. Further, prompt dissemination and data sharing is of critical importance to ensure consistently high *external reliability* across groups and experimental setups.

- *Validity* is difficult to assess using noninvasive neuroimaging methodologies. Intracranial methods (animal models, human intracranial EEG) can provide ground truth assessment to improve evaluation of biomarker *validity*. Furthermore, simultaneous scalp and intracranial EEG may allow for direct comparisons between noninvasive biomarkers and their neural correlates (12). We strongly recommend that investigators initiate collaborations with labs using these intracranial methods to assess biomarker *validity*.

**Figures and Legends**

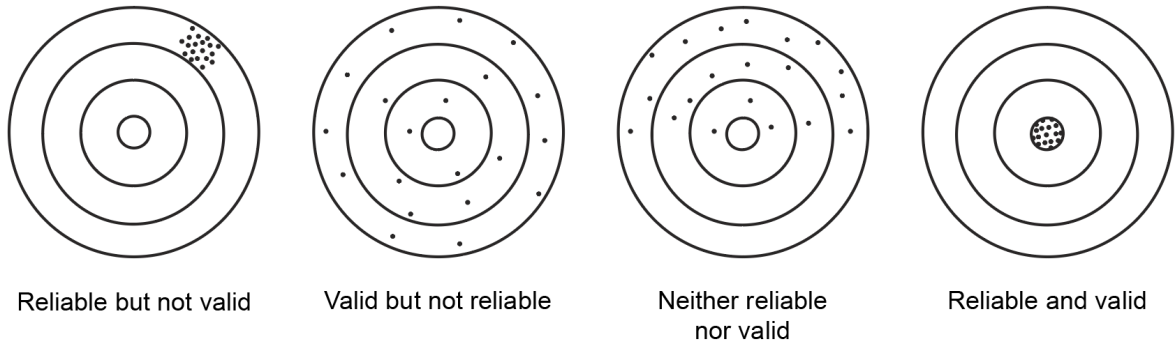

Reliable but not valid    Valid but not reliable    Neither reliable nor valid    Reliable and valid

**Figure 1 - Reliability and validity of neuroimaging biomarkers.** A biomarker should be scrutinized for both reliability and validity because it is possible to have high reliability but low validity or high validity but low reliability.

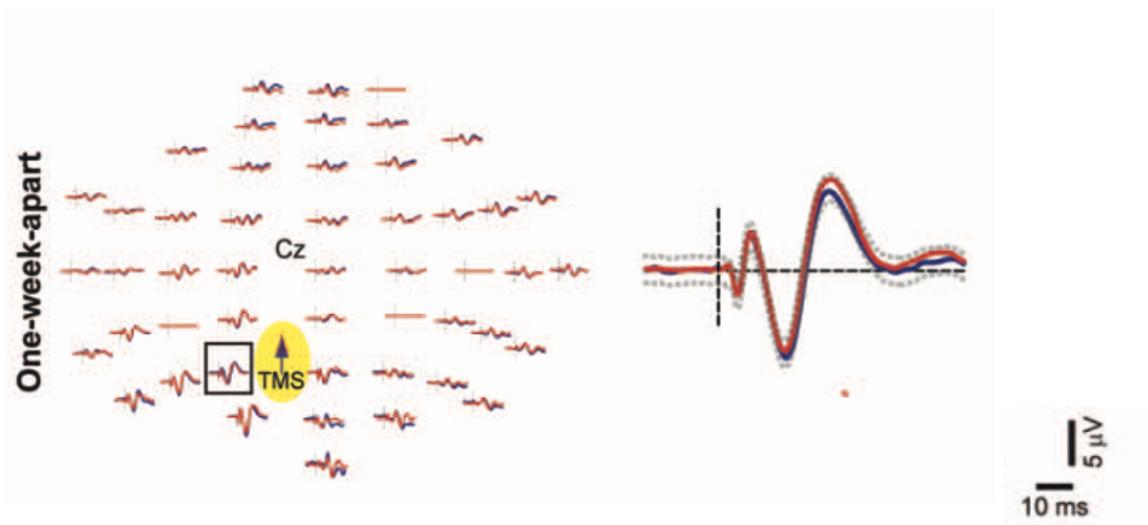

**Figure 2 - High *internal reliability* of TEPs.** One week apart, TEPs resulting from TMS to the same stimulation site on the same individual were found to be consistent. Vertical dotted line denotes timing of TMS pulse. Adapted from (29).

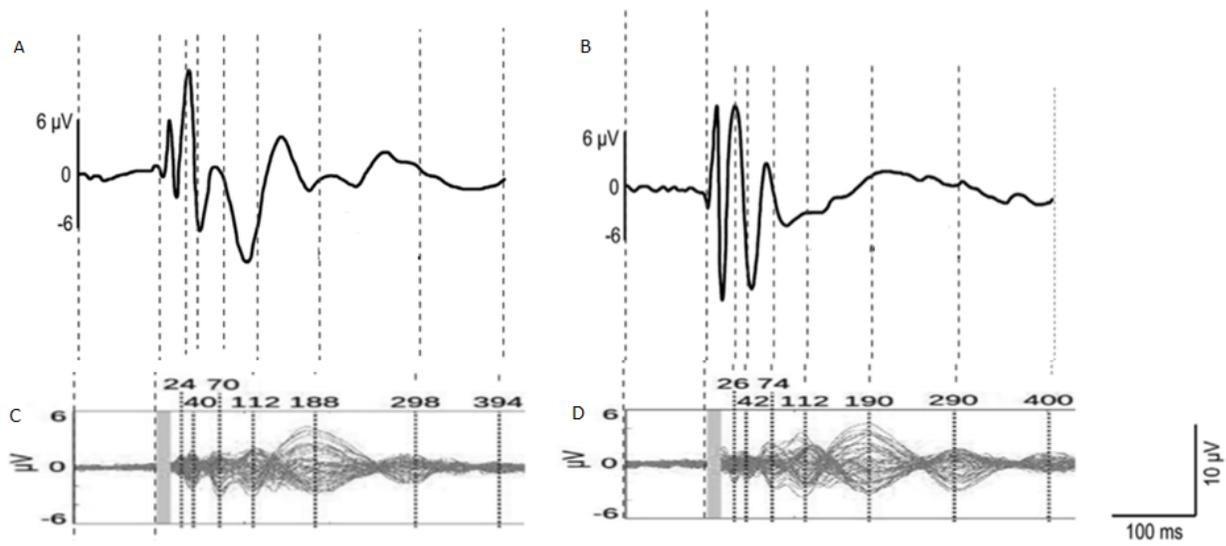

**Figure 3 -** **Low *external reliability* of the TEP**. Comparison of two TEPs resulting from different setups with stimulation of the same brain areas. Local TEP responses after stimulation of parietal (A-C) and frontal cortex (B-D). The top row of TEPs are from (37) (A-B), and the bottom row from (5) (C-D). Adapted from (46).

**Disclosures**

xDr. Parmigiani, Ross, Cline, and Minasi report no biomedical financial interests or potential conflicts of interest. Dr. Gogulski receives funding from Orion Research Foundation and Finnish Medical Foundation. Dr. Keller receives funding from the National Institutes of Health, the Burroughs Wellcome Fund, and holds equity in Alto Neuroscience, Inc.